\documentclass[journal]{IEEEtran}

\usepackage{graphicx}
\usepackage{color}
\usepackage[numbers, square, comma, sort&compress]{natbib}
\usepackage{amsmath,amsthm}
\usepackage{amssymb, bm, dsfont}
\usepackage{float}
\usepackage{tikz}
\usepackage{pgfplots}
\pgfplotsset{compat=newest} 
\pgfplotsset{plot coordinates/math parser=false}
\usepackage{amssymb,stmaryrd,amsmath,amsfonts,rotating}
\usepackage{booktabs}

\begin{document}

\title{Design Aspects of Multi-Soliton Pulses for Optical Fiber Transmission}
%
%
%


%
\author{\IEEEauthorblockN{Vahid Aref\IEEEauthorrefmark{1},
Zhenhua Dong\IEEEauthorrefmark{2}, 
and 
Henning Buelow\IEEEauthorrefmark{1}
}
\\
\IEEEauthorblockA{\IEEEauthorrefmark{1} Nokia Bell Labs, Nokia, Stuttgart, Germany, Email: first.last@nokia-bell-labs.com}
\\
\IEEEauthorblockA{\IEEEauthorrefmark{2} Photonics Research Centre, The Hong Kong Polytechnic University, Kowloon, Hong Kong}}

%
%

\maketitle

\begin{abstract}
We explain how to optimize the nonlinear spectrum of multi-soliton pulses 
by considering the practical constraints of transmitter, receiver, and lumped-amplified link.
The optimization is applied for the experimental transmission of 2ns soliton pulses with independent on-off keying of 10 eigenvalues over 2000 km of NZ-DSF fiber spans.
\end{abstract}


%
\IEEEpeerreviewmaketitle

\section{Introduction}
%
%
%
%
Nonlinear Frequency Division Multiplexing (NFDM) has recently been proposed to
design an optical waveform better matched to
the nonlinear propagation
in a fiber~\cite{yousefi2013information}.  
In particular, the information bits are modulated 
in 
the so-called nonlinear Fourier spectrum which is then
mapped to a pulse in time domain~\citep{yousefi2013information,prilepsky2014nonlinear}.
The spectrum is partitioned into continuous part and discrete part. 
  
The discrete spectrum consists of a finite set of 
complex values, called eigenvalues, and 
the corresponding 
spectral amplitudes. 
This part represents the solitonic component of the pulse, in which
the effects of Kerr nonlinearity and chromatic dispersion are
balanced.

In an ideal lossless fiber, modeled by nonlinear Schr\"odinger Equation (NLSE),
the propagation of soliton pulses follows simple principles in the nonlinear Fourier spectrum: 
the eigenvalues remain the same,
and each spectral amplitude transforms linearly only based on its eigenvalue. 
This suggests to modulate or/and detect data over nonlinear spectrum.

The first realization of such a modulation was the 
on-off keying of first-order soliton 
which has been well-studied in the last three decades, see \cite{mollenauer2006solitons} and references therein. 
However, the spectral efficiency can be increased
by using soliton pulses with several eigenvalues:    
On-off keying of up to 4 eigenvalues, located on imaginary axis, was experimentally shown~\cite{dong2015nonlinear} as well as the QPSK-modulation of spectral amplitudes for 2-soliton pulses~\citep{aref2015experimental,aref2016design}. More recently,
two of the authors showed the transmission of soliton pulses with seven eigenvalues 
and QPSK-modulated spectral amplitudes~\cite{Buelow20167eigenvalues}.

In this paper, we briefly explain how to optimize 
the nonlinear spectrum of a multi-soliton pulse 
in order to reduce the perturbations caused by the practical constraints of transmitter, link and receiver.
Applying the optimization, we demonstrate the experimental transmission of 2ns multi-soliton pulses 
carrying 10 information bits by on-off keying of 10 eigenvalues over 2033 km of NZ-DSF fiber spans.  

\section{Optimization of Multi-soliton Pulses}

\begin{figure}[bh!]
   \centering
   \input{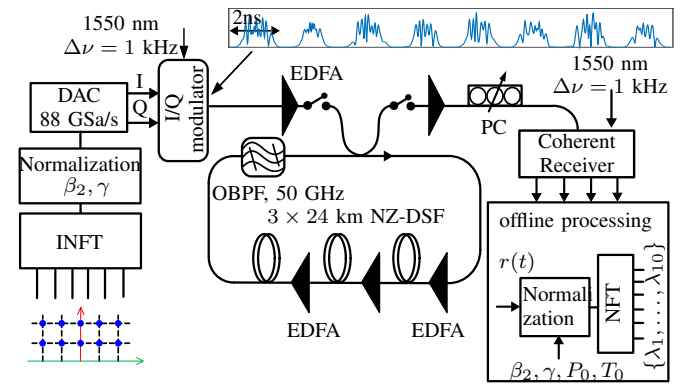}
    \caption{\label{fig:setup}Experimental setup with offline NFT-based detection}
    
\end{figure}
\begin{figure*}[tb!]
\vspace*{-2cm}
\newlength{\picwidth}
\newlength{\picheight}
\newlength{\temp}
\setlength{\picwidth}{0.27\textwidth}
\setlength{\picheight}{0.17\textwidth}
  \centering
\input{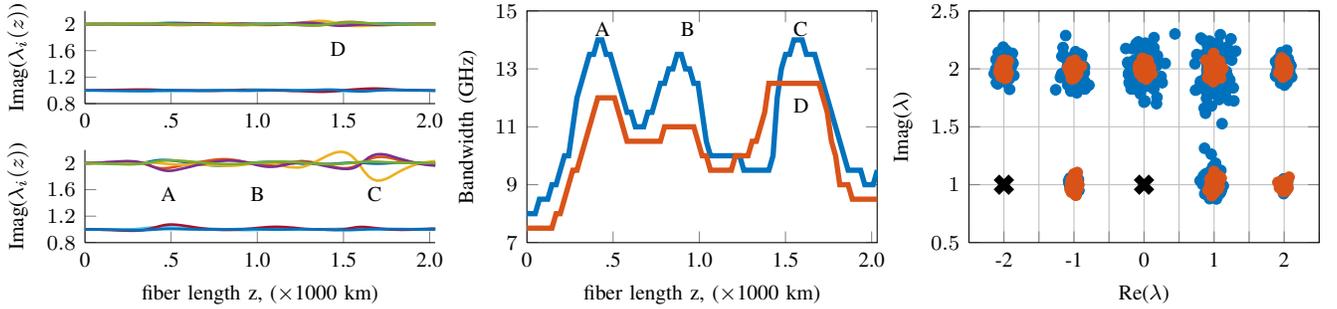}
  \caption{Simulated transmission of two soliton pulses with the same 8 eigenvalues but different initial spectral amplitudes over an EDFA amplified NZ-DSF link. Left: In the absence of noise, perturbations of imaginary part of eigenvalues for both pulses  
Middle: The bandwidth variation of pulses and its relation to perturbations of eigenvalues. Right: In the presence of noise, eigenvalues of the received pulses for both solitons after 2000 km with OSNR$\approx 34 dB$.}
\label{fig:pulseshape}
\vspace*{-.1cm}
\end{figure*}
We explain our sub-optimal method for the on-off keying modulation in which the spectral amplitudes can be freely tailored to meet the physical constraints. Consider the on-off keying of up to 10 eigenvalues for a target transmission length of at most 2100 km. Our objective is to optimize 
the discrete spectrum such that
the following conditions are satisfied for
all $2^{10}$ possible soliton pulses:

$(i)$ Pairwise distance between eigenvalues is large enough for reliable detection in the presence of 
perturbations from lumped amplification~\cite{aref2016design}, ASE noise, 
and numerical limitation of current NFT algorithms.

$(ii)$ The largest pulse-width (duration) of all pulses
must become minimum (highest transmission rate).
We truncate the tails of each pulse outside a \textit{common} interval $T$. To have a negligible inter-symbol interference (ISI), we choose $T$ such that
$|s(t)/\sqrt{E_s}|<0.01$ 
for $t\notin T$ and for all solitons $s(t)$ with energy $E_s$.

$(iii)$ The bandwidth of mulit-solitons is changing during the transmission. The largest bandwidth (BW) of all pulses must become minimum all over the transmission (highest spectral efficiency). We define bandwidth as the frequency range holding more than $99\%$ of energy.

$(iv)$ A further ISI may occur when $\text{Re}\{\lambda_i\}\ne 0$. There must be a negligible ISI between adjacent pulses during the transmission.

It is not yet fully understood which distribution of eigenvalues maximizes the spectral efficiency. This is because no analytic expression is yet available for bandwidth and pulse-width of multi-solitons and for perturbations of eigenvalues by noise and the lumped amplification. We consider here the simplest (but not the most efficient) distribution for eigenvalues: $\lambda=\omega+j\sigma$, with $\omega=\{\pm 2,\pm 1,0\}$ and $\sigma=\{1,2\}$. 

The magnitude of spectral amplitudes, $q_d(\lambda)$, mainly controls the pulse-width and the phase of 
$q_d(\lambda)$ controls the bandwidth. 
To avoid ISI (condition $(iv)$), we set $|q_d(\lambda_i)|=A_d(\lambda_i) |\exp(+2j\lambda_i^2z_L)|$, where $z_L$ is the maximum link length (2000 km) and $A_d(\lambda_i)$ is $|q_d(\lambda_i)|$  in $z_L/2$. In this case, a pulse will first be contracted and then eventually broadened to the same pulse-width it had at transmitter. To minimize the pulse-width, we optimize $A_d(\lambda_i)$. We numerically observe that
the pulse-width becomes minimum if $|B_i|=|A_i|=1$ in Darboux transformation for computing inverse nonlinear Fourier transform~\citep{yousefi2013information,aref2016control}. Finally, we computed the pulse-width $T=12$ (for standard nonlinear Schr\"odinger equation) which scales down to $2$ ns in our transmission setup ($5$ Gbit/s).

The bandwidth (BW) of each soliton is important not only because of spectral efficiency, but also because of \textit{deterministic} perturbations of eigenvalues in a lumped amplified link. It is shown in \cite{aref2016design} that the eigenvalues of the path-averaged solitons may fluctuate when their instantaneous BW are large. The same effect is shown here in Fig.~\ref{fig:pulseshape} for two soliton pulse with the same 8 eigenvalues but different spectral amplitudes. Split-step Fourier method is used to simulate the pulse propagation in our experimental setup (Fig.~\ref{fig:setup}) with NZ-DSF fiber with $\beta_2\approx-5.75 {\rm ps^2/km}$ and $\gamma\approx1.6 {\rm W^{-1}/km}$ and span length of 24.2 km. We first exclude noise in our simulation. Fig.~\ref{fig:pulseshape} shows that the eigenvalues fluctuate when the BW gets large. The larger BW is, the larger fluctuations of eigenvalues are.
In the presence of noise (assumed 50 GHz filtered noise with NF=5 dB), the soliton with larger eigenvalue fluctuations is more distorted.
For each soliton, we should then find the spectral amplitudes which minimize the maximum BW over the link. We quatized the phase of $q_d$ to levels of $\pi/4$ and used the exhaustive search to find the soliton with minimum BW for all spans.

\section{Experimental results and Conclusion}

\begin{figure}[tb!]
  \centering
\input{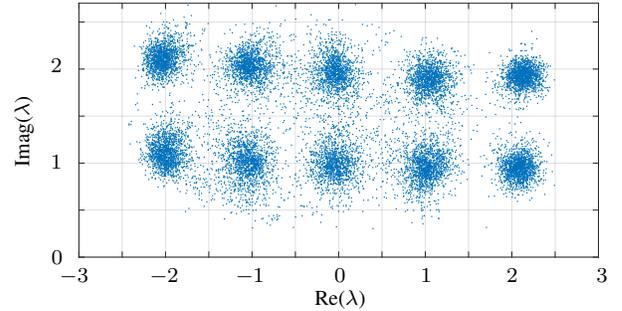}
  \caption{The eigenvalues of received pulses after 2033 km.}
\label{fig:experiment_detection}
\vspace*{-.1cm}
\end{figure}

The experimental setup is shown in Fig.~\ref{fig:setup}. Following a 88GSa/s digital-to-analog converter (DAC), a drive signal is provided for a Mach-Zehnder IQ modulator which transmits a single polarization 0.5 GBd stream into a link of up to 28 loops ($3\times L_{\rm span}$= 72.6 km) of NZ-DSF fiber with a mean launch power of about $-2.7{\rm dBm}$. 
We used homodyne detection with a low phase noise fiber laser (1 kHz linewidth). The received signal is coherently detected and sampled by an oscilloscope with 80GSa/s and 33 GHz bandwidth, followed by an offline data-aided phase and carrier offset correction. 

A subset of $2^8$ solitons were randomly chosen such that each eigenvalue is ``on'' in one-half of pulses, and a ``fair'' number of $k-$soliton pulses are chosen, $1\leq k\leq 10$.
The pulse train is repetitively transmitted.
We used Fourier Collocation (FC) method~~\cite{yousefi2013information} 
to detect the eigenvalues of each received pulse.
The eigenvalues of all received pulses after 28 loops ($\approx 2033$ km) 
are plotted in Fig.~\ref{fig:experiment_detection}. By mapping each pulse to 10 bits, 
we found the BER$\approx 8\times 10^{-3}$.

In this paper, we briefly explained how the eigenvalues of a multi-soliton are sensitive to the BW in a periodically lumped amplified link and how to optimize a soliton pulse using the available degrees of freedom in spectral amplitudes.

\ifCLASSOPTIONcaptionsoff
  \newpage
\fi



%

\bibliographystyle{IEEEtran}
\small
\bibliography{references}

\begin{thebibliography}{1}
\providecommand{\url}[1]{#1}
\csname url@samestyle\endcsname
\providecommand{\newblock}{\relax}
\providecommand{\bibinfo}[2]{#2}
\providecommand{\BIBentrySTDinterwordspacing}{\spaceskip=0pt\relax}
\providecommand{\BIBentryALTinterwordstretchfactor}{4}
\providecommand{\BIBentryALTinterwordspacing}{\spaceskip=\fontdimen2\font plus
\BIBentryALTinterwordstretchfactor\fontdimen3\font minus
  \fontdimen4\font\relax}
\providecommand{\BIBforeignlanguage}[2]{{%
\expandafter\ifx\csname l@#1\endcsname\relax
\typeout{** WARNING: IEEEtran.bst: No hyphenation pattern has been}%
\typeout{** loaded for the language `#1'. Using the pattern for}%
\typeout{** the default language instead.}%
\else
\language=\csname l@#1\endcsname
\fi
#2}}
\providecommand{\BIBdecl}{\relax}
\BIBdecl

\bibitem{yousefi2013information}
M. Yousefi, ``Information transmission using the nonlinear fourier
  transform,'' Ph.D. dissertation, University of Toronto, 2013.

\bibitem{prilepsky2014nonlinear}
J.~E. Prilepsky, \textit{et al.}
  ``Nonlinear inverse synthesis and eigenvalue division multiplexing in optical
  fiber channels,'' \emph{Phys. review letters}, vol. 113, no.~1, p. 013901,
  2014.

\bibitem{mollenauer2006solitons}
L.~F. Mollenauer and J.~P. Gordon, \emph{Solitons in optical fibers:
  fundamentals and applications}.\hskip 1em plus 0.5em minus 0.4em\relax
  Academic Press, 2006.

\bibitem{dong2015nonlinear}
Z.~Dong, \textit{et al.} ``Nonlinear frequency division multiplexed
  transmissions based on nft,'' \emph{IEEE PTL},
  vol.~27, no.~15, 2015.

\bibitem{aref2015experimental}
V.~Aref, \textit{et al.} ``Experimental demonstration of
  nonlinear frequency division multiplexed transmission,'' in Proc. ECOC, 2015.

\bibitem{aref2016design}
V.~Aref, H.~Buelow, ``Design of 2-soliton spectral phase modulated pulses
  over lumped amplified link,'' in Proc. ECOC, 2016.

\bibitem{Buelow20167eigenvalues}
H.~Buelow, \textit{et al.} ``Transmission of waveforms determined by 7
  eigenvalues with psk-modulated spectral amplitudes,'' in Proc. ECOC, 2016.

\bibitem{aref2016control}
V.~Aref, ``Control and detection of discrete spectral amplitudes in nonlinear
  fourier spectrum,'' \emph{arXiv preprint:1605.06328}, 2016.

\end{thebibliography}

%

%
%
%




\end{document}